\documentclass[twocolumn,english,showpacs,preprintnumbers]{revtex4}
\usepackage{ae,aecompl}
\usepackage[T1]{fontenc}
\usepackage[latin9]{inputenc}
\usepackage{units}
\usepackage{amsmath}
\usepackage{graphicx}

\makeatletter
\@ifundefined{textcolor}{}
{%
 \definecolor{BLACK}{gray}{0}
 \definecolor{WHITE}{gray}{1}
 \definecolor{RED}{rgb}{1,0,0}
 \definecolor{GREEN}{rgb}{0,1,0}
 \definecolor{BLUE}{rgb}{0,0,1}
 \definecolor{CYAN}{cmyk}{1,0,0,0}
 \definecolor{MAGENTA}{cmyk}{0,1,0,0}
 \definecolor{YELLOW}{cmyk}{0,0,1,0}
}

\usepackage{dcolumn}

\usepackage{bm}

\makeatother

\usepackage{babel}

\makeatother

\usepackage{babel}

\makeatother

\usepackage{babel}
\begin{document}

\title{Production of bright entangled photons from moving optical boundaries}

\author{Ariel Guerreiro}

\email{ariel@fc.up.pt}

\affiliation{INESC - Porto, Rua do Campo Alegre, 687, Porto, Portugal}

\affiliation{Departamento de F\'{\i}sica, Faculdade Ciências Universidade do Porto,
687 4169-007, Porto, Portugal.}

\author{Aires Ferreira}

\affiliation{CFP and Departamento de F\'{\i}sica e Astronomia, Faculdade Ciências
Universidade do Porto, 687 4169-007, Porto, Portugal.}

\author{J. T. Mendonça}

\affiliation{IPFN, Instituto Superior Técnico, Av. Rovisco Pais 1, 1000-69 Lisboa,
Portugal}

\date{\today}
\begin{abstract}
We discuss a mechanism of generating two separable beams of light
with high degree of entanglement in momentum using a fast and sharp
optical boundary. Three regimes of light generation are identified
depending on the number of resonant interactions between the optical
perturbation and the electromagnetic field. The intensity of the process
is discussed in terms of the relevant physical parameters: variation
of refractive index and apparent velocity of the optical boundary.
Our results suggest a different class of generation entangled light
robust against thermal degradation by exciting zero point fluctuations
using parametric resonant optical modulations. 
\end{abstract}
\maketitle
\emph{Introduction}. Many of the theoretical schemes and experimental
applications being proposed and developed in the context of Quantum
Information (QI) (including quantum computation and information processing
\cite{QC}, teleportation \cite{Teleportation}, etc.) rely on the
generation of entanglement between different quantum systems. Though
entanglement can arise in nature even from the simplest interactions
and even at high temperature \cite{Aires,Vlatko}, the degree of entanglement
achieved is usually very small. An important exception are photons,
which combined with their resilience to thermal effects, can be used,
for example, to establish quantum communication at long-distances
\cite{Zeilinger}. Till now, entangled photons are produced experimentally
via parametric down conversion (PDC), which is in general a nonlinear
process with small efficiency \cite{Boyd,Valencia}.

This Rapid Communication is motivated by the need of sources of photonic
entanglement with finer brightness and improved contrast \cite{PDCHighlyEfficient,Braunstein}.
In our proposal, high quality two-photon entangled states are spontaneously
emitted out of the vacuum (or a thermal state) by a superluminal modulation
of the refractive index of an optical medium, such as a semiconductor
where the sudden creation of electron-hole pairs can reduce the refractive
index from $\sim3.5$ to almost $0$ \cite{Yablonovitch}, or a gas
sweeped by a laser or electron beam and producing a plasma via photoionization
\cite{Oliveira,Fisher,Lampe}. Recently, a gaussian beam was sent
into a plasma inducing a superluminal two-photon ionization fronts
and used for optical-to-Thz photon conversion \cite{Experiments,Kostin}.
We show that similar techniques can generate highly-entangled photons
with a mean number of pairs that can be made arbitrarily high by increasing
the sharpness of the induced refractive index variation and by tuning
the apparent velocity of the optical modulation and the phase velocity
of the electromagnetic modes (superluminal resonance). For current
state-of-art experimental values, our estimates suggest that it is
possible to produce photons in excess of $10^{10}s^{-1}$. The main
limitation comes from the difficulty in producing an optical modulation
close enough to the resonance conditions. These results open doors
to the efficient generation of entangled photons with very high signal-to-noise
ratio via time-dependent optical perturbations, and to potential application
for QI and quantum metrology experiments.

\emph{Optical moving boundaries}. Recently, a series of papers \cite{Time Refraction,Mendoncabook,Review}
introduced the concept of Time Refraction (TR) to describe how the
classical and quantum properties of light are altered by the sudden
change of the optical properties of a medium. TR results from the
symmetry between space and time, extending the usual concept of refraction
into the time domain. Like the Unruh effect \cite{Unruh}, the Hawking
mechanism \cite{Hawking} and the dynamical Casimir effect \cite{Schultzhold},
the quantum theory of TR predicts the excitation of virtual particle
from the turmoil of Zero-Point Fluctuations (ZPF) and the emission
of pairs of real counter-propagating photons, which (as we will show)
are higly entangled. The number of pairs emitted is proportional to
the variation of the refractive index of light associated with the
optical perturbation. For any realistic experimental parameters, the
mean photon number produced in the optical domain from the vacuum
state is smaller than $1$. To overcome this limitation, a different
process of excitation of ZPF was proposed in a recent work \cite{Guerreiro2005},
using a non-accelerated optical boundary moving with $\textit{apparent}$
superluminal velocity across an optical medium. Like TR, this effect
also leads to the emission of photons pairs, but now the moving optical
boundary works as a relativistic partial mirror, producing a considerable
Doppler shift, altering radically the intensity of the interaction
between light and matter, yielding a potentially measurable number
of photons by choosing adequately the velocity of the optical boundary. 

Extending the results in \cite{Guerreiro2005} from a one dimensional
to a three dimensional geometry, we consider an infinite optical medium
swept by an optical perturbation, described as a sharp variation of
the refractive index of the medium with \textit{apparent} superluminal
velocity $\mathbf{u}$ (see Figure 1). In this context, the apparent
velocity $\mathbf{u}$ describes a delay of the change of refractive
index between different points of space and does not refer to an actual
velocity of propagation of the optical profile. Hence $u\equiv|\mathbf{u}|$
can take values arbitrarily large, even larger than $c$.

We describe the process of interaction between the ZPF and the optical
perturbation in a reference frame $S^{\prime}$, with velocity $v_{\infty}\equiv-c^{2}/u<c$
relative to the laboratory reference frame $S$, where this optical
boundary is perceived as moving with a infinite velocity: $u^{\prime}=\underset{v\rightarrow v_{\infty}}{\lim}\left(u+v\right)/\left(1+(vu)/c^{2}\right)\rightarrow\infty.$
As a consequence of the relativistic phase invariance, the refractive
index of the medium in the $S$ frame and in the $S^{\prime}$ frame
(respectively $n$ and $n^{\prime}$) are different \cite{Mendoncabook}
\begin{equation}
n^{\prime}=n\frac{\left[\gamma^{2}(\cos\theta-\beta/n)^{2}+\sin^{2}\theta\right]^{1/2}}{\gamma^{2}(1-\beta n\cos\theta)},\label{eq21}
\end{equation}
 where $\beta\equiv v/c=-c/u$, $\gamma=\left(1-\beta^{2}\right)^{-1/2}$
and $\theta$ is the angle between the velocity $\mathbf{u}$ and
the wave vector $\mathbf{k}$.

In the $S^{\prime}$ frame the problem is identical to a TR and can
be solved by imposing the continuity of the dielectric displacement
and the magnetic induction fields and corresponding field operators
\cite{Cirone} during the time discontinuity of the refractive index,
or equivalently, by imposing phase matching conditions at the optical
boundary. Back in the $S$ frame, the optical perturbation can be
perceived as a four-port device, coupling two initial complex plane
wave modes: $\phi_{i}(\mathbf{r})=\exp[-i\mathbf{k}_{i}\cdot\mathbf{r}]$
and $\phi_{a}(\vec{r})=\exp[-i\mathbf{k}_{a}\cdot\mathbf{r}]$ existing
for $\mathbf{r}>\mathbf{u}t$, with two final complex plane wave modes
$\phi_{t}(\mathbf{r})=\exp[-i\mathbf{k}_{t}\cdot\mathbf{r}]$ and
$\phi_{r}(\mathbf{r})=\exp[-i\mathbf{k}_{r}\cdot\mathbf{r}]$ existing
for $\mathbf{r}<\mathbf{u}t$, which satisfy

\begin{eqnarray}
\mathbf{k}_{t} & = & k_{i}\gamma^{2}\left[f+\sigma_{it}g_{t}\right]\mathbf{u}_{\parallel}+k_{i}\sin\theta_{i}\mathbf{u}_{\perp}\label{eq. doppler 1}\\
\mathbf{k}_{r} & = & -k_{i}\gamma^{2}\left[f-\sigma_{ir}g_{r}\right]\mathbf{u}_{\parallel}-k_{i}\sin\theta_{i}\mathbf{u}_{\perp}\label{eq. doppler 2}\\
\mathbf{k}_{a} & = & -k_{i}\gamma^{2}\left[f+g_{i}\right]\mathbf{u}_{\parallel}-k_{i}\sin\theta_{i}\mathbf{u}_{\perp},\label{eq. doppler 3}
\end{eqnarray}
 where $f\equiv(\cos\theta_{i}-\beta/n_{i})$, $\sigma_{it}\equiv h_{i}/h_{t}$,
$\sigma_{ir}\equiv h_{i}/h_{r}$, $h_{i,t,r}\equiv\left[\gamma^{2}(\cos\theta_{i}-\beta/n_{i,t,r})^{2}+\sin^{2}\theta_{i}\right]^{1/2}$,
$g_{t,r}\equiv\beta(1-\beta n_{t,r}\cos\theta_{i})/n_{t,r}$, $\theta_{i}$
now is the angle between the velocity of the optical perturbation
$\mathbf{u}$ and the wave vector $\mathbf{k}_{i}$. The different
values of the refractive index $n_{i}$, $n_{t}$, $n_{r}$ and $n_{a}$
for the incident, transmitted, reflected and anti-incident waves take
into account the dispersion of the medium prior and after optical
perturbation has passed. 
\begin{figure}
\begin{centering}
\includegraphics[width=0.5\columnwidth]{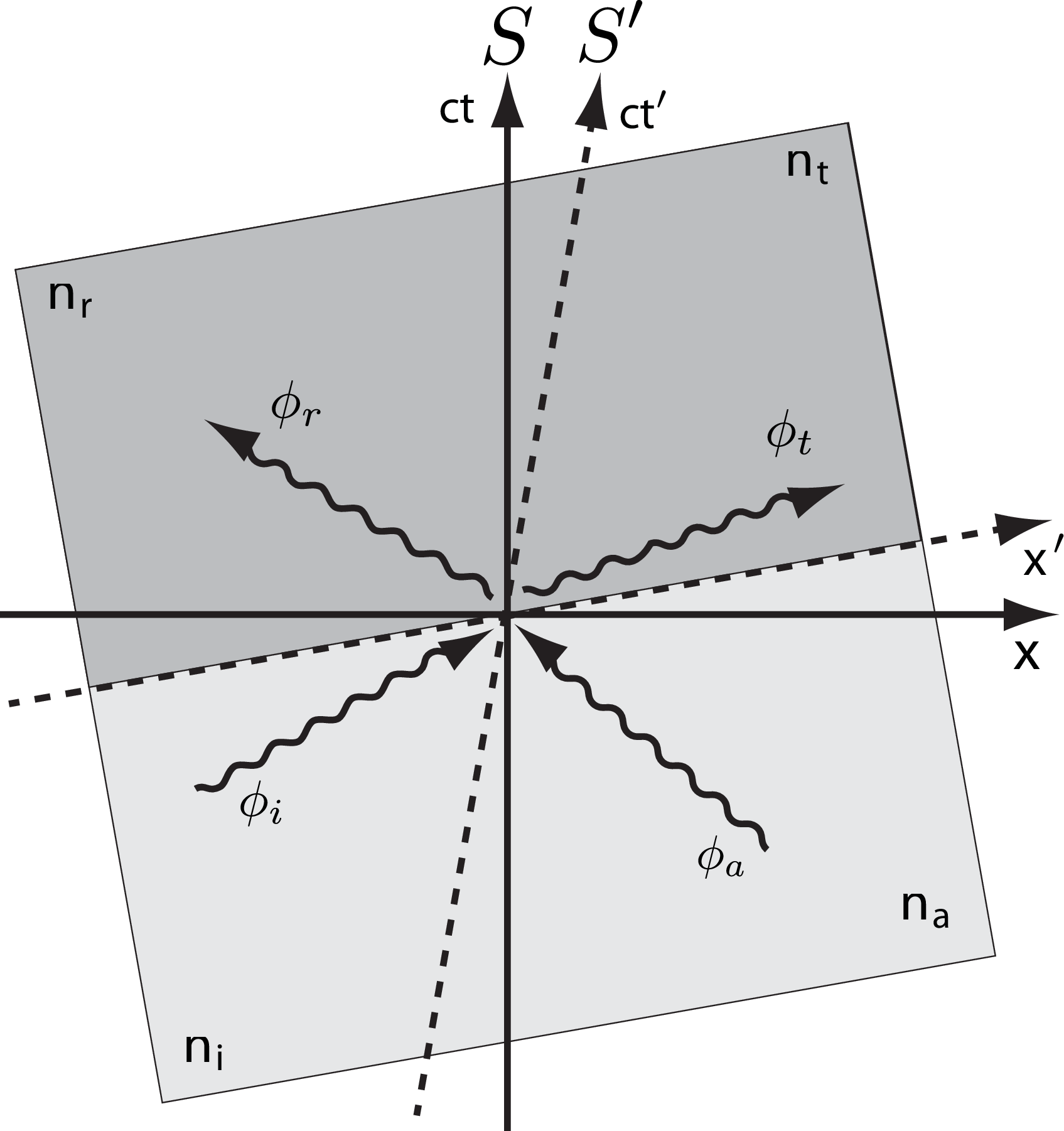} 
\par\end{centering}

\caption{Space-time schematic diagram of superluminal space-time refraction.
In the $S$ frame (bold), the optical perturbation is observed as
moving along the $x$ axis (from left to right) with apparent velocity
$u$ whereas, in the $S'$ frame (dashed), the optical perturbation
alters the refractive index of the medium simultaneously for all point
of space at instant $t'=0$. Both frames can be related using a standard
Lorentz boost with $\beta=v_{\infty}/c=-c/u<1$.}
\end{figure}

Like Eq. (\ref{eq21}), Eqs. (\ref{eq. doppler 1}) to (\ref{eq. doppler 3})
are also derived from the invariance of the phase of light between
any two different inertial frames \cite{Mendoncabook} and correspond
to a double Doppler shift. For values $\theta_{i}\neq0,\pi$, $\phi_{t}$
and $\phi_{r}$ are calculated as
\begin{eqnarray}
\theta_{t} & = & \arctan\left[\sin\theta_{i}/\gamma^{2}\left(f+\sigma_{it}g_{t}\right)\right],\label{eq teta t}\\
\theta_{r} & = & \arctan\left[\sin\theta_{i}/\gamma^{2}\left(f-\sigma_{ir}g_{r}\right)\right].\label{eq teta r}
\end{eqnarray}
 These expressions correspond to the generalized Fresnel formula for
a moving superluminal partial mirror.

Using the continuity conditions for dielectric displacement and the
magnetic induction fields \cite{Cirone,Time Refraction} at time $\mathbf{r}=t\mathbf{u}$,
the annihilation and creation operators for these modes can be related
as
\begin{equation}
a_{i}=Aa_{t}-Ba_{r}^{\dagger}\text{ and }a_{a}=Aa_{r}-Ba_{t}^{\dagger},\label{squeezing}
\end{equation}
 where $A=\left(1+\alpha^{2}\right)/2\alpha$, $B=\left(1-\alpha^{2}\right)/2\alpha$
and $\alpha=\left[n_{i}^{2}g_{i}h_{i}/n_{t}^{2}g_{t}h_{t}\right]^{1/2}$,
satisfying $A^{2}-B^{2}=1$.

As demonstrated in Refs. \cite{Leonhardt,Gilles}, the two-mode squeezing
transformation (\ref{squeezing}) implies that, after the optical
perturbation has passed, an initial vacuum can be expressed in terms
of the new eigenstates of the field as 
\begin{equation}
\left\vert 0\right\rangle _{i}\left\vert 0\right\rangle _{a}=\sum{}_{n}C_{n}\left\vert n\right\rangle _{t}\left\vert n\right\rangle _{r},\label{eq36}
\end{equation}
 with $C_{n}=\sqrt{1-\left\vert z\right\vert ^{2}}z^{n}$ and $z=B/A$.
Eq. (\ref{eq36}) implies the emission of photon pairs moving along
the different directions of $\mathbf{k}_{t}$\ and $\mathbf{k}_{r}$,
according to Eqs. (\ref{eq. doppler 1}) to (\ref{eq. doppler 3}).
The mean photon number for wave vectors $\mathbf{k}_{t}$\ and $\mathbf{k}_{r}$
is
\begin{equation}
\left\langle N_{t}\right\rangle =\left\langle N_{r}\right\rangle =\frac{|z|^{2}}{1-|z|^{2}}=\frac{\left[n_{i}^{2}h_{i}g_{i}-n_{t}^{2}h_{t}g_{t}\right]^{2}}{4n_{i}^{2}n_{t}^{2}h_{i}h_{t}g_{i}g_{t}}.\label{photon number}
\end{equation}

\begin{figure}
\begin{centering}
\includegraphics[width=0.7\columnwidth]{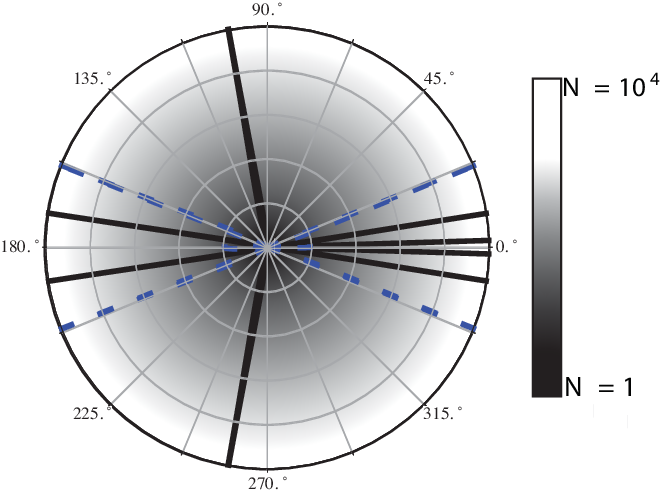} 
\par\end{centering}

\caption{\textbf{Emission spectrum:} Angular distribution of the number of
photons emitted by the optical boundary corresponding to a change
of refractive index from $n_{i}=n_{a}=1.1$ to $n_{t}=n_{r}=1.5$
and for $\beta=0.9$ (dashed) and for $\beta=0.99$ (bold). Notice
that the emission predominantly concentrated in a very small solid
angle around the resonance angles. With $\beta=0.9$ there is only
the resonance for $\beta n_{i}\cos\theta_{i}^{res}\rightarrow1$,
whereas with $\beta=0.99$, both resonances $\beta n_{i}\cos\theta_{i}^{res}\rightarrow1$
and $\beta n_{r}\cos\theta_{r}^{res}\rightarrow1$ exist. }
\end{figure}

According to Eq. (\ref{photon number}) the number of photons emitted
diverges for $z\rightarrow1$. In the one-dimensional case studied
in \cite{Guerreiro2005} this could only be achieved if a perfect
matching between the velocity of the optical perturbation $u$ and
$n_{i}$ such that $\beta n_{i}=cn_{i}/u\rightarrow1$. However, in
the three-dimensional case there is an extra degree of freedom corresponding
to the angle between the wave vector and the direction of the apparent
motion of the optical perturbation, and the condition $z\rightarrow1$
can be achieved for both $\beta n_{i}\cos\theta_{i}^{res}\rightarrow1$
and/or $\beta n_{r}\cos\theta_{r}^{res}\rightarrow1$. Unlike the
case of TR, the photon emission produced by a superluminal optical
perturbation is not limited by the maximum variation of refractive
index produced by optical perturbation. Instead, when the phase velocity
of the waves $\phi_{i}$ and $\phi_{r}$ along $\mathbf{u}$ are identical
to $\mathbf{u}$, corresponding respectively to $\beta n_{i}\cos\theta_{i}^{res}\rightarrow1$
and/or $\beta n_{r}\cos\theta_{r}^{res}\rightarrow1$, the optical
perturbation and the waves $\phi_{i}$ and $\phi_{r}$ move together
and can interact for longer times producing an arbitrarily large number
of photons. This process can be described as a form of superluminal
resonance. We identify three regimes: i) for $\beta n_{i}<1$ and
$\beta n_{r}<1$ there are no resonances; ii) for either $\beta n_{i}>1$
or $\beta n_{r}>1$ there is only one pair of resonant emission angles;
and iii) for both $\beta n_{i}>1$ or $\beta n_{r}>1$ two pairs of
resonant light are emitted. An extra resonance also exists for $\beta=n$
in media such as plasmas, where the refractive index is lower than
$1$ \cite{Guerreiro2005}, for simplicity we neglect this resonance
herein. These resonances can be achieved for a wide range of experimental
parameters and configurations. The angular distribution corresponding
to Eq. (\ref{photon number}) is represented in Figure 2 where we
can clearly identify $\theta_{t}^{res}$ and $\theta_{r}^{res}$,
calculated from $\theta_{i}^{res}$ using Eqs. (\ref{eq teta t})
and (\ref{eq teta r}) respectively. Notice that emission is mainly
limited to a narrow solid angle, resulting in colimated beams.

\emph{Photonic entanglement generation}. The field can be separated
into two subsystems ($S$ and $S^{\prime}$) corresponding to the
two distinct sets of photons emitted, i.e. $\phi_{r}$ and $\phi_{t}$.
Depending on the initial state of the field, these two subsystems
may become entangled after the optical perturbation. We discuss and
compare the degree of entanglement between two situation: an initial
vacuum and a thermal state (which is the experimental case).

According to Eq. (\ref{squeezing}), an initial vacuum state is changed
into another pure state for which the entanglement entropy $E_{VN}(\rho_{SS^{\prime}})\equiv-\textrm{Tr}\left[\rho^{S}\ln\rho^{S}\right]$
(with $\rho^{S}=Tr_{S^{\prime}}\left[\rho_{SS^{\prime}}\right]$)
is the canonical entanglement measure \cite{Shumacher}, yielding
$E_{VN}=\ln\left(1+\left\langle N_{t}\right\rangle \right)\left(1+\left\langle N_{t}\right\rangle \right)-\left\langle N_{t}\right\rangle \ln\left\langle N_{t}\right\rangle $.
Notice that $E_{VN}$ is basically the Shannon entropy introduced
by increasing a photon pair in the system. For $z\rightarrow1$, the
entanglement diverges as the system approaches the resonance condition
and the maximal entanglement state is achieved, i.e.
\begin{equation}
\underset{z\rightarrow1}{\lim}\left\vert 0\right\rangle _{i}\left\vert 0\right\rangle _{a}=\underset{N\rightarrow\infty}{\lim}\underset{n=0}{\overset{N}{\sum}}\frac{1}{\sqrt{N}}\left\vert n\right\rangle _{t}\left\vert n\right\rangle _{r}.
\end{equation}

If the system is initially in a thermal state of both wave modes,
$\phi_{i}$ and $\phi_{a}$, i.e. $\rho_{ia}(\bar{n})=\rho(\bar{n})\otimes\rho(\bar{n})$,
with 
\begin{equation}
\rho(\bar{n})=\frac{1}{1+\bar{n}}\sum_{n=0}^{\infty}\left(\frac{\bar{n}}{1+\bar{n}}\right)^{n}|n\rangle\langle n|,\label{eq:ThermalState-1}
\end{equation}
where $\bar{n}$ is the thermal mean occupancy, then after the optical
perturbation has passed, the state $\rho_{tr}(z)$ describing the
$\phi_{t}$ and $\phi_{r}$ modes becomes $\rho_{tr}(z)=S(z)\rho_{ia}(\bar{n})S(z)^{\dagger}$,
which is a squeezed thermal state \cite{Marian} and for which $E_{VN}$
is not an adequate entanglement measure \cite{Plenio}. However $\rho_{tr}$
is a Gaussian state, and its entanglement can be completely characterized
using continuous variable methods (see \cite{Ferraro} for a review),
namely via the \emph{logarithmic negativity}, $E_{N}(\rho)=\max\left[0,-\ln\mu\right]$,
where $\mu$ is the smallest sympletic eigenvalue of the Gaussian
state $\rho_{tr}$. The expression for $\mu$ (see \cite{Laurat}
for a derivation) is $\mu=\left(2\bar{n}+1\right)\exp\left(-2\text{arctanh}z\right)$.
The latter defines a thermal occupancy $n_{c}$, above which all entanglement
vanishes, yielding $2\bar{n}_{c}+1=\exp\left(2\text{arctanh}z\right).$
Close to resonance ($z\rightarrow1$) the maximum allowable thermal
occupancy diverges $\bar{n}_{c}\rightarrow\infty$; entailing that
entanglement extraction from optical boundaries is very robust regarding
temperature by choosing a sufficiently high $z$.

\emph{Discussion of efficiency}. Now we consider an optical perturbation
in a frame $S$ of the form, $n(x-ut)=n_{0}+\delta n\ f[K(x-ut)]$,
where $K$ is a spatial scale describing the sharpness and duration
of the optical perturbation and $f(x)=0$ for $x\leq0$, $f(x)=x$
for $0<x\leq1$ and $f(x)=1$ for $x>1$. In the $S'$ frame, the
creation and annihilation operators in the interaction picture satisfy
\cite{Tito2005}
\begin{equation}
\frac{d}{dt'}a_{t}=\nu(t')a_{r}^{\dagger},\ \frac{d}{dt'}a_{r}^{\dagger}=\nu^{*}(t')a_{t},
\end{equation}
with $\nu(t')=\nicefrac{1}{2}exp[2i\psi(t')][\nicefrac{d}{dt'}ln\ n']$,
where $\psi(t')$ is a phase. The total photon number $N=N_{t}+N_{r}$
satisfies
\begin{equation}
\frac{d^{2}}{dt'^{2}}(N+2)=4\left|\nu(t')\right|^{2}(N+2).
\end{equation}
For a small variation of refractive index (i.e. $\delta n\ll n_{0}$),
the total number of photons produced, the maximum and average rate
of photon generation from initial thermal states are respectively
\begin{eqnarray}
N_{total} & \approx & (N_{0}+2)\ \cosh(\eta\delta n)\\
R_{max} & \equiv & \frac{dN}{dt}\approx(N_{0}+2)\eta\delta nKu\ \sinh(\eta\delta n)\\
R_{mean} & \approx & (N_{0}+2)Ku\gamma^{-1}\ \cosh(\eta\delta n)
\end{eqnarray}
where $\eta\equiv\nicefrac{d}{dn_{0}}ln\ n_{0}'$. For conditions
close the resonances, $\eta\sim1/(n_{0}\Delta)$, where $\Delta\equiv1-\beta n\cos\theta$
is the detuning from the resonance conditions. 

\emph{Conclusions}. We presented an emission mechanism of entangled
radiation using a sharp optical perturbation with an apparent superluminal
velocity. The emission spectrum and the emissivity depend on the apparent
velocity and the change of refractive index of the optical perturbation.
These results extend those of reference \cite{Guerreiro2005} from
a one-dimensional configuration to one that includes all complex plane-wave
modes in a three-dimensional space and is valid for an arbitrary dispersive
medium. For our particular configuration, the optimum direction of
emission is defined by the resonances $\Delta_{i}\rightarrow0$ and
$\Delta_{r}\rightarrow0$. The resonance angles $\theta_{t}^{res}$
and $\theta_{r}^{res}$ correspond to both the best radiance and to
the optimally entangled photons. From a purely theoretical point of
view, this process has considerable advantages over PDC as a source
of entangled light, namely, since it is capable of delivering two
well separable and highly entangled beams with large intensities.
In our case the photons are entangled in momentum whereas in PDC the
photons are entangled in polarization; however, these two types of
entanglement can be interconverted \cite{Boschi}. From a more experimental
point of view, it is not easy to produce a sharp and sudden optical
perturbation at scales inferior to the optical wavelengths to allow
the large number of photon pairs necessary to make this process competitive
with PDC. A conservative estimate based on parameters from present
day experimental demonstrations of superluminal ionization fronts\cite{Experiments,Kostin,TeraHz Pulses}
($\beta\approx0.9995$, $K/c\approx0.02fs$, assuming $\delta n/n_{0}\approx1\%$)
predicts photon yields in excess of $R_{max}\sim10^{10}s^{-1}$ ($R_{mean}\sim10^{9}s^{-1}$)
for $\Delta\sim0.01$. Moreover, a recent work has shown that this
quantum mechanism of extracting photon pairs out of ZPF can be extended
to optical perturbations with arbitrary shape as long as they have
an \textit{apparent} superluminal velocity \cite{Tito2005}. These
results suggest the possibility of generating high intensity entangled
photons via specific time-dependent optical perturbations, including
dynamical Casimir effect.

A. G. acknowledges the support of the Casimir network of the European
Science Foundation. A.F. acknowledges the support of FCT (Portugal)
through grant PRAXIS no. SFRH/BD/18292/04.

\end{document}